\documentclass[letterpaper, 10pt, conference]{ieeeconf}

\IEEEoverridecommandlockouts

\usepackage{amsmath,mathtools}  %
\usepackage{amsfonts,amssymb}

\usepackage{xspace}    %
\usepackage{balance}

\usepackage{graphicx}
\usepackage{subcaption}
\graphicspath{{figs/}}

\usepackage{pgfplots}
\usepackage{grffile}
\pgfplotsset{compat=newest}  %
\usetikzlibrary{plotmarks}
\usetikzlibrary{arrows.meta}
\usepgfplotslibrary{patchplots}

\usepackage{mathrsfs}
\usetikzlibrary{arrows}
\usepackage{tabularx}

\usepackage[ruled]{algorithm}
\usepackage[noend]{algpseudocode}

\makeatletter
\newcommand{\multiline}[1]{%
  \begin{tabularx}{\dimexpr\linewidth-\ALG@thistlm}[t]{@{}X@{}}
    #1
  \end{tabularx}
}
\makeatother

\usepackage[per-mode=symbol]{siunitx}

\usepackage{cite}  %
\usepackage{array}  %
\usepackage{booktabs}           %

\usepackage{cite}

\newtheorem{assumption}{Assumption}

\newcounter{remark}
{\par\endtrivlist\unskip}

\newcounter{problem}
{\par\endtrivlist\unskip}

\usepackage[extramath,probability,reviewmark]{truonglatexdefs}

\newcommand{\GPModel}{\ensuremath{\GGG}}

\newcommand{\GPmean}{\ensuremath{\mu}}

\renewcommand{\IdentityMatrix}{\mathbb{I}}  %

\title{A Receding Horizon Approach for \\ Simultaneous Active Learning and Control using Gaussian Processes}

\author{Viet-Anh Le and Truong X. Nghiem\\
  School of Informatics, Computing, and Cyber Systems\\
  Northern Arizona University\\
  \{vl385,truong.nghiem\}@nau.edu%
}

\begin{document}

\maketitle
\thispagestyle{empty}
\pagestyle{empty}

\begin{abstract}
This paper proposes a receding horizon active learning and control problem for dynamical systems in which Gaussian Processes (GPs) are utilized to model the system dynamics.
The active learning objective in the optimization problem is presented by the exact conditional differential entropy of GP predictions at multiple steps ahead, which is equivalent to the log determinant of the GP posterior covariance matrix.
The resulting non-convex and complex optimization problem is solved by the Sequential Convex Programming algorithm that exploits the first-order approximations of non-convex functions.
Simulation results of an autonomous racing car example verify that using the proposed method can significantly improve data quality for model learning while solving time is highly promising for real-time applications. 
\end{abstract}

\section{Inroduction}
\label{sec:intro}

Modeling the system dynamics play a pivotal role in the performance of model-based control techniques such as Receding Horizon Control (RHC, also known as Model Predictive Control).
Nevertheless, for many complex dynamical systems, obtained mathematical models are often insufficiently accurate due to the existence of uncertainties and ignored dynamical parts.
This challenge motivates learning-based models for control, which leverages Machine Learning (ML) techniques to model dynamical systems from data and certain prior knowledge. 
For example, Gaussian Processes (GPs) \cite{williams2006gaussian} have been applied for dynamics and control recently \cite{kocijan2016modelling}. %
A fundamental but challenging problems in learning the system dynamics using GPs is how to obtain a training dataset such that the learned models %
can efficiently capture the actual dynamics.
This is because, in most control applications, the experiments for data collection are limited by time, cost, or constraints of environment, whereas using only historical data is not suitable due to the lack of input excitation.
For instance, consider a \textbf{motivating example} of an autonomous racing car.
In this example, the experiments are constrained by narrow and sharp racing tracks, while the historical data obtained from manual control or simple automatic control techniques %
do not have sufficient excitation. 
A better approach would be simultaneous learning and control where the car explores the state space for learning as quickly as possible while satisfying other control objectives such as maintaining safety and tracking a racing trajectory. 
The learning objective requires that the system dynamics are sampled at the states associated with informative GP inputs.
Given current models, the goal can be achieved by driving the system to the state where the information from collected data %
can minimize prediction uncertainty at the region of interest. 
Once the system is controlled to the new state, the GP inputs and outputs are collected, then the model is retrained.
This repetitive procedure is referred to as \emph{active learning} for dynamical systems.

The general active learning framework has been applied to various domains (see \cite{aggarwal2014active} for a survey) to address the problem pertaining to optimal training data collection.
However, the active learning problem for dynamical systems has only been studied recently \cite{jain2018learning,capone2020localized,alpcan2015information,buisson2020actively}.
In \cite{jain2018learning}, a greedy scheme for choosing the next data point in Optimal Experiment Design (OED) was proposed by exploiting the maximum variance and information gain methods.
The models learned from the experiment was then used for an RHC problem.
Meanwhile, active learning using multi-step look-ahead was considered %
\cite{alpcan2015information,capone2020localized,buisson2020actively} and shows better performance than single-step approach.
Particularly, in \cite{capone2020localized}, based on the conditional differential entropy of multi-step GP predictions, the authors first determined the most informative data point within the region of interest by a greedy algorithm, then steering the system towards that state using RHC.
The paper \cite{alpcan2015information} considered an RHC problem for dual control (\ie simultaneous learning and control) in which the objective function consists of a control objective and a knowledge-gain (information-gain) objective for learning the dynamics.
The knowledge-gain objective was obtained using concepts from information theory, such as mutual information or relative entropy. 
In \cite{buisson2020actively}, an RHC problem for OED including both active learning objective represented by the differential entropy of GP predictions and dynamic constraints was formulated.
However, to limit the computational burden, only an upper or lower bound of the differential entropy in \cite{buisson2020actively} and the estimated knowledge gain in \cite{alpcan2015information} of the multi-step GP predictions was utilized.
That is, the information-gain metrics for multi-step GP predictions were replaced by the sum of individual metrics for step-wise predictions over the horizon.  
As a result, the problems can be solved in continuous domain by nonlinear programming solvers instead of grid-based methods. %
However, computation time for solving the optimization problems was not reported in those papers to judge whether the used methods are suitable for real-time control. 

In this paper, we presents a Receding Horizon Active Learning and Control (RHALC) problem for dynamical systems using the GP regressions.
The presented problem formulation covers both the dual control problem in \cite{alpcan2015information} and the problem for experiment design in \cite{buisson2020actively}.
However, instead of approximate information-gain metrics, we take the \emph{exact} conditional differential entropy of multi-step GP predictions into account. 
The resulting optimization problem involving the GP dynamics and the log determinants of posterior covariance matrices is thus non-convex and highly complex that may prevent the success of any non-linear programming solvers.
To overcome this challenge, we apply the Sequential Convex Programming (SCP) method \cite{mao2017successive} to efficiently address the problem by performing the first-order approximations of the GP means and the GP posterior covariance matrices. 
The effectiveness of the proposed method is validated by simulations of an autonomous racing car example. 
The results on trajectory tracking control performance of the racing car and prediction accuracy of the learned models show that the active learning can improve data quality for model learning in both offline learning (experiment design) and simultaneous learning and control.
In addition, the reported fast computation time of the SCP algorithm demonstrates its capability for real-time implementation.

The remainder of this paper is organized as follows. 
The RHALC problem formulation is introduced in Section~\ref{sec:problem}. Section~\ref{sec:algorithm} provides SCP algorithm, while the simulation results are reported and discussed in Section~\ref{sec:simulation}. 
Finally, Section~\ref{sec:conclusion} concludes the paper. %

\section{Problem Formulation}
\label{sec:problem}

In this section, we introduce a general Receding Horizon Active Learning and Control (RHALC) formulation in which the GPs are employed to model the system dynamics while the \emph{exact} conditional differential entropy is used as an optimization metric in active learning.

\subsection{Gaussian Process Regression }

Consider a latent function \(f: \RR^n \mapsto \RR\) %
and $N$ noisy observations \(y^{(i)}\) of it, \(y^{(i)} = f\left(\mathbf{x}^{(i)}\right) + \epsilon^{(i)}\), at inputs \(\mathbf{x}^{(i)} \in \RR^n\) and with i.i.d.\ Gaussian noise \(\epsilon^{(i)} \sim \GaussianDist{0}{\sigma_n^2}\), for $i = 1,\dots,N$.
We will use \(\mathbf{X} = [\mathbf{x}^{(1)}, \dots, \mathbf{x}^{(N)}] \in \RR^{n \times N}\) to denote the collection of all input vectors and \(\mathbf{Y} = [y^{(1)}, \dots, y^{(N)}] \in \RR^{N}\) to denote the collection of the corresponding observed outputs.
Let $\DDD = (\mathbf{X}, \mathbf{Y})$ be the set of observation data of $f$.
A GP of $f$, which will be denoted by $\GPModel_f$, is a probability distribution over all possible realizations of $f$ and can be formally defined as a collection of random variables, any finite number of which have a joint Gaussian distribution \cite{williams2006gaussian}.
It is fully specified by a covariance function $k(\mathbf{x},\mathbf{x}'; \theta)$ %
and a mean function $m(\mathbf{x}; \theta)$. %
The mean function is employed to include prior knowledge about the unknown function. 
In this paper, the mean function is assumed to be zero without loss of generality.

At $M$ new inputs $\mathbf{x}_\star = [\mathbf{x}_{1,\star}, \dots, \mathbf{x}_{M,\star}]$, the joint predictions at $\mathbf{x}_\star$, \(f_\star = f(\mathbf{x}_{\star})\), of $\GPModel_f$ is a random variable \(f_{\star} \sim \GaussianDist{\mu_\star}{\mathbf{\Sigma}_\star}\), in which the predictive mean vector $\mathbf{\mu}_\star$ and the $M\times M$ posterior covariance matrix $\mathbf{\Sigma}_\star$ are computed as follows 
\begin{subequations}
\begin{align}
\mathbf{\mu}_\star &= \mathbf{\mu}_{\GPModel_f}(\mathbf{x}_{\star}) = \mathbf{K}_\star (\mathbf{K} + \sigma_n^2 \II)^{-1} \mathbf{Y} \label{eq:gp-regression:mean} \\
\mathbf{\Sigma}_\star &= \mathbf{\Sigma}_{\GPModel_f} (\mathbf{x}_\star) = \mathbf{K}_{\star \star} - \mathbf{K}_\star (\mathbf{K} + \sigma_n^2 \II)^{-1}\mathbf{K}_\star^T \label{eq:gp-regression:var}
\end{align}
\end{subequations}
where $\sigma_n^2$ is the Gaussian noise variance, $\IdentityMatrix$ is an identity matrix of appropriate dimensions, $\mathbf{K}_{\star \star} \in \RR^{M\times M}$ is the covariance matrix at $\mathbf{x}_\star$, $\mathbf{K}_\star \in \RR^{M\times N}$ is the cross-covariance matrix between $\mathbf{x}_\star$ and $\mathbf{X}$, $\mathbf{K} \in \RR^{N \times N}$ is the covariance matrix at $\mathbf{X}$, in which the elements $K_{ij}$ of each matrix is computed by \(K_{ij} = k(\mathbf{x}^{(i)}, \mathbf{x}^{(j)})\). 
Note that in this paper, we only utilize the GP means without uncertainty propagation to represent the predicted values of the nonlinear dynamics.
For more details on GPs and its usage in controls, readers are referred to \cite{williams2006gaussian,kocijan2016modelling}.

\subsection{Receding Horizon Active Learning and Control with Gaussian Process}

We define the control input vector as $\mathbf{u} \in \RR^{n_{u}}$, the vector of GP output variables $\mathbf{y} \in \RR^{n_{y}}$ and the vector of non-GP variables $\mathbf{z} \in \RR^{n_{z}}$.
For any variable $\square$, where $\square$ is $\mathbf{y}$, $\mathbf{z}$, or $\mathbf{u}$, let $\square_{k}$ denote its value at time step $k$.
The state of the system at time step $k$ comprises $\mathbf{y}_{k}$ and $\mathbf{z}_{k}$.
We denote the GP dynamic model as $\mathbf{y}_{k} \sim \GPModel(\mathbf{x}_{k})$.
The input vector $\mathbf{x}_{k}$, or features, of the GP is formed from current and past values of the control inputs $\mathbf{u}_{\tau}$ and non-GP states $\mathbf{z}_{\tau}$, for $\tau \leq k$, as well as from past GP outputs $\mathbf{y}_{\tau}$, for $\tau < k$.
To simplify the notation and formulation, we assume that $\mathbf{y}$ is scalar; however, our results can be readily extended to multivariate $\mathbf{y}$.
Given an input $\mathbf{x}_{k}$, let $\mathbf{\bar{y}}_{k} = \GPmean(\mathbf{x}_{k})$ denote the predictive mean of the GP model $\GPModel(\cdot)$ at $\mathbf{x}_{k}$.

Let $H > 0$ be the length of the control horizon, $t$ be the current time step and $\III_t = \{ t, \dots, t+H-1 \}$ be the set of all time steps in the control horizon at time step $t$.
Denote $\bar{\YYY}_{ t} = \{ \mathbf{\bar{y}}_{k} | k \in \III_t \}$, $\ZZZ_{t} = \{ \mathbf{z}_{k} | k \in \III_t \}$, $\UUU_{t} = \{ \mathbf{u}_{k} | k \in \III_t \}$, and $\XXX_{t} = \{ \mathbf{x}_{k} | k \in \III_t \}$ as the sets collecting the predictive GP output means, the non-GP states, the control inputs, and GP inputs over the control horizon.
To simplify the mathematical notations, we will use $[X]$ to denote the vector concatenation of all vectors in a set $X$ (\eg $[\XXX_t] = [\mathbf{x}_{k}^T]_{k \in \III_t}^T$).

Given a GP model trained on the data generated up to the current time step, the most informative GP regressor vectors in the next horizon $\XXX_t$ can be determined by maximizing the conditional differential entropy of GP predictions at these vectors that is computed by the log determinant of the covariance matrix \cite{nguyen2015information},
leading to the following optimization problem
\begin{equation}
\XXX_t = \underset{\{\XXX_t\}}{\operatorname*{argmax}} \; \operatorname{{log\,det}} \big( \mathbf{\Sigma} ([\XXX_t]) \big) 
\end{equation}
where $\mathbf{\Sigma} ([\XXX_t])$ is a $H \times H$ posterior covariance matrix of GP predictions at $H$ input vectors in the set $\XXX_t$ and can be computed by \eqref{eq:gp-regression:var}.

The RHALC problem is thus formulated as follows
\begin{equation}
  \label{eq:rhalc}
  \begin{split}
    & \underset{\{\UUU_t, \ZZZ_t\}}{\operatorname*{minimize}} \quad J(\bar{\YYY_t}, \UUU_t, \ZZZ_t) - \gamma \HHH(\XXX_t) \\
    & \text{subject to} \\
    & \quad \bar{\mathbf{y}}_{i,k} = \mu ( \mathbf{x}_{i,k}), \; \forall k\in\III_{t}\\
    & \quad g_j(\bar{\YYY}_t, \UUU_t, \ZZZ_t) \le 0, \; \forall j \in \JJJ_{\text{ieq}}\\
    & \quad h_j(\bar{\YYY}_t, \UUU_t, \ZZZ_t) = 0, \; \forall j \in \JJJ_{\text{eq}}
  \end{split}
\end{equation} 
where $\HHH(\XXX_t) = \operatorname{{log\,det}} \big( \mathbf{\Sigma}([\XXX_t]) \big)$ is an active learning term and $J(\bar{\YYY_t}, \UUU_t, \ZZZ_t)$ is an control objective function, $g_j(\bar{\YYY}_t, \UUU_t, \ZZZ_t) \le 0$ and $h_j(\bar{\YYY}_t, \UUU_t, \ZZZ_t) = 0$ are inequality and equality constraints while $\JJJ_{\text{ieq}}$ and $\JJJ_{\text{eq}}$ are the sets of inequality and equality constraint indices, respectively.
In the problem \eqref{eq:rhalc}, $\gamma$ is a positive constant representing a tradeoff between learning and control objectives.

Similar to \cite{nghiem2019linearized}, we make the following assumption about the problem \eqref{eq:rhalc}.
\begin{assumption}
\label{assp:convexity}
Suppose that $J$ is convex, each $g_j$ is convex, and each $h_i$ is affine in the optimization variables $\UUU_t$ and $\ZZZ_t$.
In other words, the non-convexity of the problem \eqref{eq:rhalc} results from the GP dynamics and the log determinant of the GP predictive covariance matrix.
\end{assumption}

\section{Sequential Convex Programming for Receding Horizon Active Learning and Control}
\label{sec:algorithm}

The problem \eqref{eq:rhalc} is highly nonconvex due to the active learning objective and the GP dynamics.
Moreover, the complexity of the objective function involving the log determinant of the GP predictive covariance matrix makes the problem \eqref{eq:rhalc} computationally intractable. 
In this section, we employ the Successive Convex Programming (SCP) approach \cite{mao2017successive} to effectively address this problem.

Suppose that nominal feasible control inputs are given in $\UUU_t^{\star} = \{ u^{\star}_{k} \,|\, k \in \III_t \}$.
We then simulate the GP model $\GPModel(\cdot)$ over the RHC horizon to obtain the nominal output means $\overbar{\YYY}_t^{\star} = \{\overbar{y}_k^{\star} \,|\, k \in \III_t\}$.
The nominal regressor vectors in $\XXX_t^{\star} = \{ x^{\star}_{k} \,|\, k \in \III_t \}$, can be obtained from these values. %
Consider small perturbations to the nominal control inputs $u_k = u_k^{\star} + \Delta u_k$, which are collected in $\Delta \UUU_t = \{ \Delta u_k \,|\, k \in \III_t\}$.
They will cause perturbations to the predictive output means and regressor vectors during the MPC horizon as: $\Delta \overbar{\YYY}_t = \{\Delta \overbar{y}_k = \overbar{y}_k - \overbar{y}_k^{\star} \,|\, k \in \III_t\}$ and $\Delta \XXX_{y,t} = \{\Delta x_{k} = x_{k} - x_{k}^{\star} \,|\, k \in \III_t\}$.
Using these perturbation variables, the RHALC~\eqref{eq:rhalc} can be approximated locally around the nominal values by replacing the GP predictive means $\mathbf{\bar{y}}_{k} = \GPmean(\mathbf{x}_{k})$ and the log determinant of GP predictive covariance matrix with their first order approximations as follows.

Define $\tilde{y}_k$ as the first-order approximation of $\overbar{y}_k$ around the nominal solution $\mathbf{x}_k^{\star}$, which can be computed as follows
\begin{equation}
\label{eq:lingp_mean}
\tilde{y}_k = \mu (\mathbf{x}_k^{\star}) + \nabla \mu (\mathbf{x}_k^{\star})^T \mathbf{\Delta x}_k
\end{equation}
where from \eqref{eq:gp-regression:mean} we have
\begin{equation*}
\begin{split}
\mu (\mathbf{x}_k^{\star}) &= k(\mathbf{x}_k^{\star}, \mathbf{X}) (\mathbf{K}+\sigma_n^2 \II)^{-1} \mathbf{Y} \\
\nabla \mu (\mathbf{x}_k^{\star}) &= K^{(1,0)}(\mathbf{x}_k^{\star}, \mathbf{X})^T (\mathbf{K}+\sigma_n^2 \II)^{-1} \mathbf{Y}
\end{split}
\end{equation*}
with $\mathbf{K}^{(1,0)} = (\nabla_{x} k)$ being the gradient of $k$ with respect to the first argument. 
We also define $\tilde{\YYY}_t$ as a collection of $\tilde{y}_k$ for $k \in \III_t$.
Meanwhile, the first-order approximation of $\HHH(\XXX_t)$ %
around a nominal solution $[\XXX_t^{\star}]$ is computed by:
\begin{equation}
\label{eq:lin-logdet}
\begin{multlined}
\tilde{\HHH} (\Delta \XXX_t) = \operatorname{{log\,det}} \left( \mathbf{\Sigma}([\XXX_t^{\star}]) \right) \\
+ \nabla \operatorname{{log\,det}} \left( \mathbf{\Sigma}([\XXX_t^{\star}]) \right)^T [\Delta \XXX_t]
\end{multlined}
\end{equation}  

Note that the derivative of the log determinant of the GP predictive covariance matrix %
with respect to each element $\nu_j$ of a vector $\boldsymbol{\nu}$ can be computed by 
\begin{equation*}
\frac{\partial \operatorname{{log\,det}} \left( \mathbf{\Sigma}(\boldsymbol{\nu}) \right)}{\partial \nu_j} 
= \operatorname{tr} \left(\mathbf{\Sigma}^{-1} (\boldsymbol{\nu}) \frac{\partial \mathbf{\Sigma} (\boldsymbol{\nu})}{\partial \nu_j}\right)
\end{equation*} 
where from \eqref{eq:gp-regression:var} we have
\begin{equation*}
\frac{\mathbf{\Sigma} (\boldsymbol{\nu})}{\partial \nu_j} = \frac{\partial \mathbf{K}_{\star \star}}{\partial \nu_j} - 2 \mathbf{K}_\star (\mathbf{K} + \sigma_n^2 \II)^{-1} \left(\frac{\partial \mathbf{K}_\star}{\partial \nu_j}\right)^T
\end{equation*} 

With a slight abuse of notations, we will write $\UUU_t = \UUU_t^{\star} + \Delta \UUU_t$.
We now obtain the convexified RHALC problem, as stated below.
\begin{subequations}
\label{eq:lin-rhalc}
\begin{align}
   & \underset{\Delta \UUU_{t}, \ZZZ_{t}}{\text{minimize}} \;
     \; J(\tilde{\YYY}_{t}, \UUU^{\star}_{t} + \Delta\UUU_{t}, \ZZZ_{t}) 
     - \gamma \tilde{\HHH} (\Delta \XXX_t) \label{eq:lin-rhalc:obj}\\
  & \text{subject to} \nonumber \\
  & \; \tilde{y}_k = \mu (\mathbf{x}_k^{\star}) + \nabla \mu (\mathbf{x}_k^{\star})^T \mathbf{\Delta x}_k, \; \forall k\in\III_{t} \label{eq:lin-rhalc:lingp}\\  %
  & \; \norm{\Delta u_{k}}_{\infty} \leq \rho,\; \norm{\Delta x_{k}}_{\infty} \leq \rho, \; \forall k\in\III_{t} \label{eq:lin-rhalc:trustregion} \\
  & \; g_{j} (\tilde{\YYY}_{t}, \UUU^{\star}_{t} + \Delta\UUU_{t}, \ZZZ_{t}) \leq 0, \, \forall j \in \JJJ_{\text{ieq}} \label{eq:lin-rhalc:ineqs}\\
  & \; h_{j} (\tilde{\YYY}_{t}, \UUU^{\star}_{t} + \Delta\UUU_{t}, \ZZZ_{t}) = 0, \, \forall j \in \JJJ_{\text{eq}} \label{eq:lin-rhalc:eqs}
\end{align}
\end{subequations}

Constraints \eqref{eq:lin-rhalc:trustregion} specify a trust region %
in which the local convexified subproblem is valid.
To avoid the \textit{artificial infeasibility} \cite{mao2017successive} of the problem due to the approximations \eqref{eq:lingp_mean} and \eqref{eq:lin-logdet}, the inequality and equality constraints in \eqref{eq:lin-rhalc} are encoded into the objective function by the exact penalty functions \cite{mao2017successive} leading to the following penalized convex problem %
\begin{subequations}
\label{eq:pen-lin-rhalc}
\begin{align}
   & \begin{multlined}
	\underset{\Delta \UUU_{t}, \ZZZ_{t}}{\text{minimize}} \;
    \; J(\tilde{\YYY}_{t}, \UUU^{\star}_{t} + \Delta\UUU_{t}, \ZZZ_{t}) - \gamma \tilde{\HHH} (\Delta \XXX_t) \\
    + \sum_{j \in \JJJ_{\text{ieq}}}
    \tau_j \operatorname{max} \big( 0, g_{j} (\tilde{\YYY}_{t}, \UUU^{\star}_{t} + \Delta\UUU_{t}, \ZZZ_{t}) \big) \\
    + \sum_{j \in \JJJ_{\text{eq}}}
    \lambda_j \big| h_{j} (\tilde{\YYY}_{t}, \UUU^{\star}_{t} + \Delta\UUU_{t}, \ZZZ_{t}) \big|
	\end{multlined} \\
  & \text{subject to} \nonumber \\
  & \; \tilde{y}_k = \mu (\mathbf{x}_k^{\star}) + \nabla \mu (\mathbf{x}_k^{\star})^T \mathbf{\Delta x}_k, \; \ \forall k\in\III_{t} \label{eq:pen-lin-rhalc:lingp}\\  %
  & \; \norm{\Delta u_{k}}_{\infty} \leq \rho,\; \norm{\Delta x_{k}}_{\infty} \leq \rho, \; \forall k\in\III_{t} \label{eq:pen-lin-rhalc:trustregion}
\end{align}
\end{subequations}
where $\tau_j,\; \forall j \in \JJJ_{\text{ieq}}$ and $\lambda_j,\; \forall j \in \JJJ_{\text{eq}}$ are large positive penalty weights.
Under Assumption~\ref{assp:convexity}, the penalized subproblem \eqref{eq:pen-lin-rhalc} is convex and can be solved efficiently by convex solvers.
The SCP algorithm for solving the RHALC problem is outlined in Algorithm~\ref{alg:scp-rhalc}. %
For further details on the SCP method and its application to the RHC problem using the GPs, the readers are referred to \cite{mao2017successive} and \cite{nghiem2019linearized}, respectively.

\begin{algorithm}[!tb]
\caption{Successive Convex Programming for RHALC} 
\label{alg:scp-rhalc}
\begin{algorithmic}[1]
  \small%
  \Require %
  $\UUU_{t}^{(0)}$,
  $\ZZZ_{t}^{(0)}$, 
  $\rho^{(0)} > 0$, $0 < r_{0} < r_{1} < r_{2} < 1$, $\beta_{\mathrm{fail}}<1$, $\beta_{\mathrm{succ}}>1$, $\epsilon > 0$, $j_{\mathrm{max}} > 0$
  \State Simulate $\GPModel$ with $\UUU_{t}^{(0)}$, obtain $\overbar{\YYY}_{t}^{(0)}$ 
  \State $\phi^{(0)} \leftarrow \phi\big(\overbar{\YYY}_{t}^{(0)}, \UUU_{t}^{(0)}, \ZZZ_{t}^{(0)}\big)$
  \For{$j = 0, \dots, j_{\mathrm{max}}-1$}
    \State Form convex subproblem \eqref{eq:pen-lin-rhalc} by using \eqref{eq:lingp_mean} and \eqref{eq:lin-logdet}
    \State Solve problem~\eqref{eq:pen-lin-rhalc} to get $\tilde{\YYY}_{t}$, $\tilde{\UUU}_{t}$, $\tilde{\ZZZ}_{t}$
    \State Simulate $\GPModel$ with $\tilde{\UUU}_{t}$ to obtain $\overbar{\YYY}_{t}$
    \State $\delta^{(j)} \leftarrow \phi^{(j)} - \phi\big(\overbar{\YYY}_{t}, \tilde{\UUU}_{t}, \tilde{\ZZZ}_{t}\big)$
    \State $\tilde{\delta}^{(j)} \leftarrow \phi^{(j)} - \tilde{\phi}\big(\tilde{\YYY}_{t}, \tilde{\UUU}_{t}, \tilde{\ZZZ}_{t}\big)$
    \If{$|\tilde{\delta}^{(j)}| \leq \epsilon$}
      \textbf{stop and return $\UUU_{t}^{(j)}$}
    \EndIf
    \State $r^{(j)} \leftarrow \delta^{(j)} / \tilde{\delta}^{(j)}$
    \If{$r^{(j)} < r_{0}$}
      \State \multiline{Keep current solution: $\UUU_{t}^{(j+1)} \leftarrow \UUU_{t}^{(j)}$, $\ZZZ_{t}^{(j+1)} \leftarrow \ZZZ_{t}^{(j)}$, $\overbar{\YYY}_{t}^{(j+1)} \leftarrow \overbar{\YYY}_{t}^{(j)}$}
      \State $\rho^{(j+1)} \leftarrow \beta_{\mathrm{fail}}\rho^{(j)}$ 
    \Else
      \State \multiline{Accept solution: $\UUU_{t}^{(j+1)} \leftarrow \tilde{\UUU}_{t}$, $\ZZZ_{t}^{(j+1)} \leftarrow \tilde{\ZZZ}_{t}$, $\overbar{\YYY}_{t}^{(j+1)} \leftarrow \overbar{\YYY}_{t}$, $\phi^{(j)} \leftarrow \phi\big(\overbar{\YYY}_{t}, \UUU_{t}, \ZZZ_{t}\big)$}
      \If{$r^{(j)} < r_{1}$} 
        $\rho^{(j+1)} \leftarrow \beta_{\mathrm{fail}}\rho^{(j)}$
      \ElsIf{$r^{(j)} < r_{2}$}
        $\rho^{(j+1)} \leftarrow \rho^{(j)}$
      \Else{}
        $\rho^{(j+1)} \leftarrow \beta_{\mathrm{succ}}\rho^{(j)}$ 
      \EndIf
    \EndIf
  \EndFor  %
  \State \textbf{return $\UUU_{t}^{(j_{\mathrm{max}})}$}  
\end{algorithmic}
\end{algorithm}
\setlength{\textfloatsep}{0.1cm}

\section{Simulation}
\label{sec:simulation}

This section validates the advantages of the RHALC problem in two scenarios, experiment design and simultaneous learning and control problem, as well as the effectiveness of the SCP algorithm in solving the problems in real-time by a numerical simulation of an autonomous racing car example.

\subsection{Autonomous racing car example}
\label{sec:simulation:example}

We revisit the example of an autonomous racing car mentioned in Section~\ref{sec:intro}.
The simulation consists of two phases: learning phase and racing phase.
During the learning phase, starting with initial GP models trained on a few historical data, the controller collects and adds new data points and retrain the GP models.
Once sufficient data for learning accurate models have been obtained, the learning phase is disabled and the obtained models are utilized to perform tracking control task in the racing phase.

Similar to \cite{le2020gaussian}, the kinematic bicycle model of the vehicle is used in the simulation while the following discrete-time dynamics with a sampling time $\Delta T > 0$ are utilized to design the control problem
\begin{equation}
\begin{alignedat}{2}
  \label{eq:ex-disc-dyn}
  {x}_{k+1} &= x_{k} + \Delta x_{k}, \quad
  && {y}_{k+1} = y_{k} + \Delta y_{k}, \\
  {\theta}_{k+1} &= \theta_{k} + \Delta \theta_{k}, \quad
  && v_{k+1} = v_{k} + \Delta T a_{k}. \\
\end{alignedat}
\end{equation}  
where $(x, y)$ is the position vector of the vehicle, %
$\theta$ is the heading angle, $v$ is the speed of the vehicle, and $a$ and $\alpha$ are respectively the linear acceleration and steering angle of the vehicle. 
The continuous-time kinematic bicycle model and parameters of the racing car can be found in \cite{le2020gaussian}.
The nonlinear components in \eqref{eq:ex-disc-dyn} are learned by three GP models:
$\Delta x_{k} \sim \GPModel_{x} ( \mathbf{x}_{p,k} )$, $\Delta y_{k} \sim  \GPModel_{y} ( \mathbf{x}_{p,k} )$, and $\Delta \theta_{k} \sim  \GPModel_{\theta} ( \mathbf{x}_{a,k} )$, in which the vectors of GP inputs %
$\mathbf{x}_{p,k} = [\cos \theta_{k}, \sin \theta_{k}, v_{k}, \alpha_{k}]^T, \; \mathbf{x}_{a,k} = [v_{k}, \alpha_{k}]^T$. 
The GP models result in the following GP dynamical equations
\begin{equation}
  \begin{split}
    \label{eq:ex-gp-dyn}
    \Delta \bar{x}_{k} &= \mu_{ x} ( \textbf{x}_{p,k} ), \;
    \Delta \bar{y}_{k} = \mu_{ y} ( \textbf{x}_{p,k} ), \;
    \Delta \bar{\theta}_{k} = \mu_{ \theta} ( \textbf{x}_{a,k} ) \text.
  \end{split}
\end{equation}

The RHALC formulation for this example is given by
\begin{subequations}
  \label{eq:ex-gpmpc}
  \begin{align}
    & \underset{\{a_{k}, \alpha_{k}\}_{k \in \III_{t}}}{\text{minimize}} \quad J 
    - \gamma \left( \HHH_{x} + \HHH_{y} + \HHH_{\theta} \right) 
    \label{eq:ex-gpmpc:obj}\\
    & \text{subject to} \nonumber \\
    & \qquad \text{\eqref{eq:ex-disc-dyn} and \eqref{eq:ex-gp-dyn}} \\
    & \qquad v_{\text{min}} \leq v_{k} \leq v_{\text{max}}, \label{eq:ex-gpmpc:velo-bound-cons}\\   
    & \qquad a_{\text{min}} \leq a_{k} \leq a_{\text{max}}, \quad \alpha_{\text{min}} \leq \alpha_{k} \leq \alpha_{\text{max}} \label{eq:ex-gpmpc:bound-cons} \\
    & \qquad \mathbf{a}_k   \begin{bmatrix}
    x_{k+1} & y_{k+1}
  \end{bmatrix}^T \le \mathbf{b}_k \label{eq:ex-gpmpc:safety-cons}
  \end{align}
\end{subequations} 
where the constraints hold for all $k \in \III_{t}$, \eqref{eq:ex-gpmpc:velo-bound-cons} are velocity bound constraints, \eqref{eq:ex-gpmpc:bound-cons} are bound constraints on the control actions, and \eqref{eq:ex-gpmpc:safety-cons} are affine constraints %
representing collision avoidance to the border of the racing track.
The objective of this problem consists of two parts: trajectory tracking control and active learning for dynamics.
The control objective function $J$ is given by
\begin{equation*}
  J = \sum_{k = t}^{t+H-1} \left\|
  \begin{bmatrix}
    a_{k} \\ \alpha_{k}
  \end{bmatrix}
  \right\|_{\mathbf{R}}^2 + 
  \left\|
  \begin{bmatrix}
    x_{k+1} \\ y_{k+1}
  \end{bmatrix} - \mathbf{r}_{k+1} \right\|_{\mathbf{Q}}^2 
\end{equation*}
where $\mathbf{r}_{k+1}$ denotes the reference at time step $k+1$.
Given a vector $\nu$ and a positive semidefinite matrix $\mathbf{M}$, we define $\norm{\nu}_{\mathbf{M}}^2 = \nu^T \mathbf{M} \nu$.  
The active learning goals for the GP models are captured by %
\begin{equation*}
\begin{split}
\HHH_{x} &= \operatorname{{log\,det}} \big(\mathbf{\Sigma}_{\GPModel_{x}} (\mathbf{x}_{p,t+1:t+H}) \big) \\
\HHH_{y} &= \operatorname{{log\,det}} \big(\mathbf{\Sigma}_{\GPModel_{y}} (\mathbf{x}_{p,t+1:t+H}) \big) \\
\HHH_{\theta} &= \operatorname{{log\,det}} \big(\mathbf{\Sigma}_{\GPModel_{\theta}} (\mathbf{x}_{a,t+1:t+H}) \big)  
\end{split}
\end{equation*}
where $\mathbf{x}_{p,t+1:t+H}$ and $\mathbf{x}_{p,t+1:t+H}$ denotes the concatenated vector of GP inputs from time $t+1$ to $t+H$.
To avoid the collision with the track borders, we employed a scheme presented in \cite{liniger2015optimization} that limit the movement of the car at each time step in the horizon to lie within two parallel half-planes. %
As a result, the collision avoidance scheme can be represented as a set of linear inequality constraints in \eqref{eq:ex-gpmpc:safety-cons}.

\begin{figure*}[!tb]
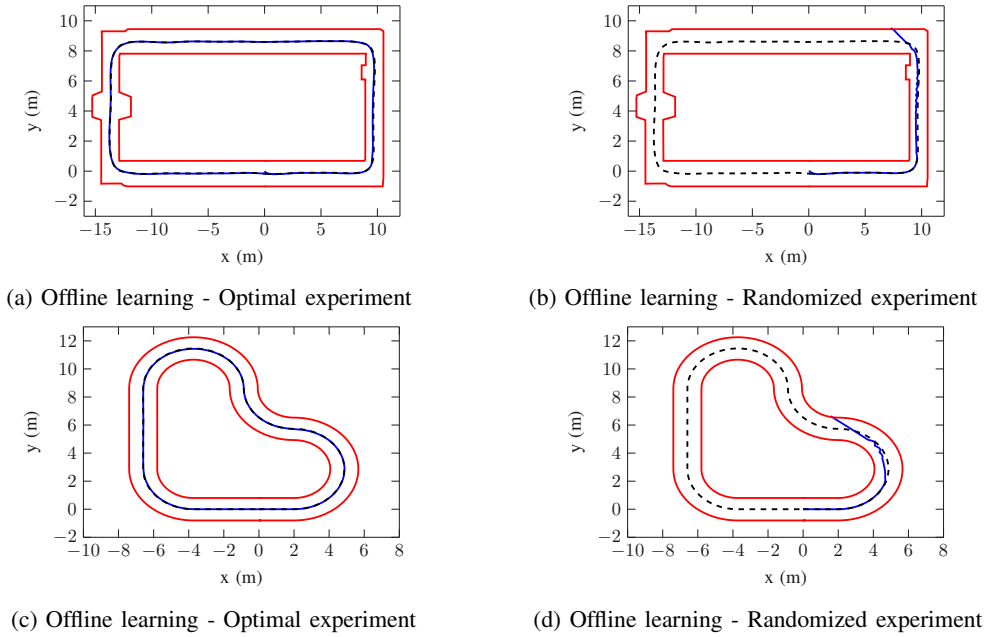

\vspace{15pt}
\centering
\begin{subfigure}{.4\textwidth}
\centering
\scalebox{0.7}{\input{figs/upenn_oed.tex}}
\caption{Offline learning - Optimal experiment}
\end{subfigure}
\begin{subfigure}{.4\textwidth}
\centering
\scalebox{0.7}{\input{figs/upenn_re.tex}}
\caption{Offline learning - Randomized experiment}
\end{subfigure}
\vspace{5pt}

\begin{subfigure}{.4\textwidth}
\centering
\scalebox{0.7}{\input{figs/ucb_oed.tex}}
\caption{Offline learning - Optimal experiment}
\end{subfigure}
\begin{subfigure}{.4\textwidth}
\centering
\scalebox{0.7}{\input{figs/ucb_re.tex}}
\caption{Offline learning - Randomized experiment}
\end{subfigure}
\caption{The trajectories (blue lines) of the autonomous vehicle in two racing tracks 
with the offline GP models from optimal experiment ((a) and (c)) and randomized experiment ((b) and (d)).}
\label{fig:offline}
\vspace{-15pt}
\end{figure*} 

The sampling time $\Delta T =$ \SI{200}{ms} 
while the control horizon length $H = 5$.
The parameters in the constraints are:
$v_{\mathrm{min}} = \SI{0}{m/s}$,
$v_{\mathrm{max}} = \SI{2}{m/s}$,
$\alpha_{\mathrm{min/max}} = \SI{\pm \pi/4}{rad/s}$,
$a_{\mathrm{min/max}} = \SI{\pm 2}{m/s^2}$.
The weights in the objective function are: 
$\mathbf{Q} = \operatorname{diag}([10^2,10^2])$,
$\mathbf{R} = \operatorname{diag}([10^{-1}, 10^{-1}])$,
$\gamma = 10$.
All the penalty weights in \eqref{eq:pen-lin-rhalc} are chosen to be $10^6$.
The parameters of the SCP algorithm are:
$\rho^{(0)} = 0.1$,
$\beta_{\mathrm{fail}} = 0.5$,
$\beta_{\mathrm{succ}} = 2.0$,
$r_0 = 0.01$,
$r_1 = 0.1$,
$r_2 = 0.3$,
$j_{\mathrm{max}} = 100$,
$\epsilon = 10^{-4}$.
Two virtual racing tracks developed by the University of California, Berkeley (UCB) \cite{rosolia2019learning} and the University of Pennsylvania (UPenn) \cite{o2020f1tenth} are used in the simulations. 

In what follows, we present two scenarios for the racing car example considered in this paper: Offline learning and Simultaneous Learning and Control (Online Learning). 

\subsubsection{Offline learning}
Assume that a large area without obstacles is available for experiments. %
We design an experiment to obtain an optimal dataset for training the GP dynamics prior to the race.
This procedure is also referred to as \emph{Optimal Experiment Design} (OED)
\cite{buisson2020actively}.
In the OED problem, only the active learning term is included in the objective function while the control objective $J$ is removed. %
The constraints \eqref{eq:ex-gpmpc:safety-cons} %
are simplified to bound constraints on $x$ and $y$ %
to ensure that the car lies within the experimental space.

We compare the optimal experiment design using RHALC with a randomized experiment design where random control inputs that satisfy all the input constraints are applied to the system. 
In the optimal experiment design, three simple GP models with 25 initial data points are utilized, then the RHALC problem is applied in 50 time steps to collect new data points. %
The models learned from experimental data are used to perform a reference tracking control task. %
The receding horizon reference tracking problem can be derived by removing the active learning term in \eqref{eq:ex-gpmpc}. %

\subsubsection{Simultaneous Learning and Control}
In the application where a free area for experiments is not available, it is required to perform \emph{Simultaneous Active Learning and Control}, or Online Learning, in the learning phase.
The vehicle is expected to efficiently collect online data to update the GP dynamics while following the racing track and avoiding collision to the borders.
Hence, this scenario covers the dual control problem in \cite{alpcan2015information}.

To validate the benefits of active learning, we consider two simulations depending on the effect of the active learning term. 
In the first simulation, the active learning objective is involved to %
whereas in the second simulation, the active learning is disabled and the vehicle only tracks to the reference and collects online data along the racing track.
In this scenrio, three initial GP models trained on 50 initial data points are adopted to control the system at the beginning, then 100 new data samples are collected online.
The simultaneous learning and control simulation is conducted using the UPenn track, while the UCB track is utilized for a testing race given the model learned from the previous race.

\subsection{Results and discussions}

\begin{figure*}[!tb]
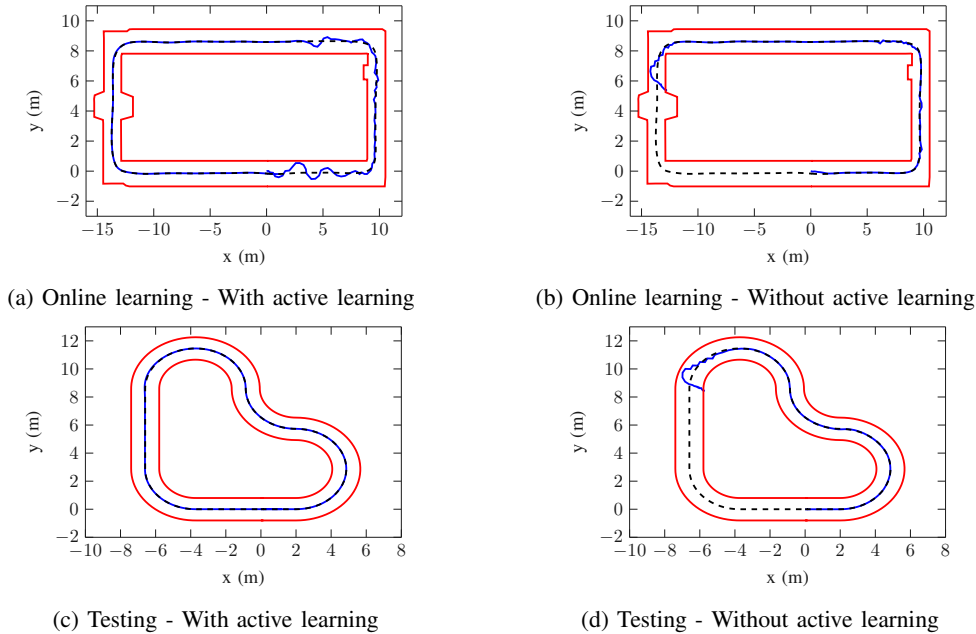

\vspace{15pt}
\centering
\begin{subfigure}{.4\textwidth}
\centering
\scalebox{0.7}{\input{figs/upenn_al.tex}}
\caption{Online learning - With active learning}
\end{subfigure}
\begin{subfigure}{.4\textwidth}
\centering
\scalebox{0.7}{\input{figs/upenn_wal.tex}}
\caption{Online learning - Without active learning}
\end{subfigure}
\vspace{5pt}

\centering
\begin{subfigure}{.4\textwidth}
\centering
\scalebox{0.7}{\input{figs/ucb_al.tex}}
\caption{Testing - With active learning}
\end{subfigure}
\begin{subfigure}{.4\textwidth}
\centering
\scalebox{0.7}{\input{figs/ucb_wal.tex}}
\caption{Testing - Without active learning}
\end{subfigure}

\caption{The trajectories (blue lines) of the autonomous vehicle in two racing tracks 
with the GP models updated online with active learning ((a) and (c)) and without active learning ((b) and (d)).}
\label{fig:online}
\vspace{-15pt}
\end{figure*} 

The trajectories of the autonomous racing car in the offline learning simulation are shown in Fig.~\ref{fig:offline}. 
Using the GP models from the optimal experiment, the receding horizon tracking controller can accurately track the reference and prevent the collision to the borders.  
In contrast, the GP models from the randomized experiment apparently are not accurate %
since the car cannot complete one racing lap without crashing into the borders. 
Particularly, the car initially can track well to the reference but collision to the borders happens when the car needs to turn sharply.  
Meanwhile, Fig.~\ref{fig:online} shows the trajectories of the car in the Simultaneous Learning and Control simulation. 
As can be seen from the figure, with the active learning, %
the vehicle initially fluctuates around the reference to explore the informative states, hence the car does not perfectly track the reference in the first 100 time instants but the safety condition is guaranteed.
However, once completing the learning phase, the obtained GP models are accurate so that the vehicle is able to track the reference trajectory in the rest of the racing track and in a new testing track.
Meanwhile, without active learning, at the beginning of the race where the racing track is relatively simple, then the car can track to the reference.
However, since the learned models do not have enough excitation, 
the car deviates to the reference trajectory when the track becomes more complicated %
leading to the crashes to the borders.  

Furthermore, the prediction accuracy of the GP models obtained from all simulations are shown in Table~\ref{tab:metrics}.
We compare 4 types of GP models, OE, RE, AL and Non-AL, which are correspondingly obtained from the offline optimal and randomized experiments, %
and the learning and control simulations with and without active learning. %
Two validation metrics including the root mean squared errors (RMSEs), and the maximum absolute errors (MAEs) are considered.
These validation metrics are computed using the GP predictions on a grid of GP inputs in which 20 linearly spaced points in the region of interest for each input
and the corresponding latent non-linear functions. %
According to the table, with the same number of training data points, three GP models generated from the optimal experiment show better performance in prediction accuracy than those from the randomized experiment. 
Likewise, based on the metrics for AL and non-AL models, it is obvious that active learning can improve GP precision in simultaneous learning and control.    

Regarding the computation time, in the optimal experiment, SCP algorithm takes $\SI{0.069}{s}$ on average for each time instant, while in the race, it takes $\SI{0.041}{s}$ and $\SI{0.040}{s}$ in UCB and UPenn racing tracks, respectively.
Meanwhile, in online learning simulation that includes active leaning, SCP algorithm averagely takes $\SI{0.104}{s}$, whereas in control phase, it takes $\SI{0.058}{s}$ for each time step.
Note that all simulations in this work are performed on a DELL computer with a \SI{3.0}{GHz} Intel Core i5 CPU and \SI{8}{Gb} RAM, and the Julia programming language is used for the implementation.

\setlength{\extrarowheight}{.1cm}
\begin{table}[!bt]
\vspace{4pt}
\caption{Validation metrics for different types of obtained GP models.}
\label{tab:metrics} 
\centering
\begin{tabular}{ c | c c | c c | c c }
\toprule[1pt]%
& \multicolumn{2}{c|}{$\GPModel_{x}$} & \multicolumn{2}{c|}{$\GPModel_{y}$} & \multicolumn{2}{c}{$\GPModel_{\theta}$} \\
& RMSE & MAE & RMSE & MAE & RMSE & MAE \\
\midrule[0.5pt] %
OE & 0.050 & 0.151 & 0.058 & 0.189 & 0.006 & 0.017 \\
RE & 0.139 & 0.443 & 0.084 & 0.397 & 0.020 & 0.046 \\
AL & 0.036 & 0.192 & 0.041 & 0.161 & 0.015 & 0.070 \\
Non-AL & 0.063 & 0.296 & 0.114 & 0.498 & 0.038 & 0.238 \\
\bottomrule[1pt] %
\end{tabular}
\end{table}

\section{Conclusion}
\label{sec:conclusion}

In this paper, a receding horizon active learning and control problem for dynamical systems using Gaussian Processes (GPs) was considered. 
We first developed a problem formulation subjecting to the GP dynamics while the exact conditional differential entropy was employed as a metric for active learning. 
The resulting complex and non-convex problem was solved by the Sequential Convex Programming algorithm. 
The proposed method was validated by numerical simulations of an autonomous racing car example.
Not only guarantee the tracking performance in both offline and online learning scenarios, the control algorithm but also can be executed in a reasonable amount of time, which promises its potential practicality for real-time implementation.

\balance

\bibliographystyle{IEEEtran}
\bibliography{IEEEabrv,references}

\end{document}